\documentstyle[prl,aps,epsfig,multicol]{revtex}

\begin{document}

\def\be{\begin{equation}}
\def\ee{\end{equation}}
\def\bearr{\begin{eqnarray}}
\def\eearr{\end{eqnarray}}
\def\tc{$T_c~$}
\def\tcl{$T_c^{1*}~$}
\def\c2{ CuO$_2~$}
\def\ruo{ RuO$_2~$}
\def\lsco{LSCO~}
\def\bi{bI-2201~}
\def\tl{Tl-2201~}
\def\hg{Hg-1201~}
\def\sro{$Sr_2 Ru O_4$~}
\def\rc{$RuSr_2Gd Cu_2 O_8$~}
\def\mgb{$MgB_2$~}
\def\pz{$p_z$~}
\def\ppi{$p\pi$~}
\def\sqo{$S(q,\omega)$~}
\def\tperp{$t_{\perp}$~}
\def\cob{$CoO_2$~}
\def\nxcob{$Na_x CoO_2.yH_2O$~}
\def\ncob{$Na_{0.5} CoO_2$~}
\def\half{$\frac{1}{2}$~}
\def\nycob{$A_xCoO_{2+\delta}$~}


\title{Chiral Spin Fluctuations in planar \cob and Possibility of \\
Chiral RVB Metal and Superconductivity}
\title{An Electronic Model for layered \nxcob:\\
 A case for Anyonic Superconducting State ?}
\title{An Electronic Model for \cob layer based systems: \\
 Chiral RVB metal and Superconductivity}

\author{ G. Baskaran\\
The Institute of Mathematical Sciences\\
C.I.T. Campus, Chennai 600 113, India }


\maketitle

\begin{abstract}

Takada et al. have reported superconductivity in layered \nxcob 
(\tc$\approx5~K$) and more recently Wen et al. in \nycob 
($A = Na,K$)(\tc$\approx~31~K$). We model a reference neutral \cob layer
as an orbitally non-degenerate spin-\half antiferromagnetic Mott insulator 
on a triangular lattice and \nxcob and \nycob as electron doped Mott 
insulators 
described by a t-J model. It is suggested that at optimal doping chiral 
spin fluctuations enhanced by the dopant dynamics leads to a 
$d$-wave superconducting state. A chiral RVB metal, a PT violating state 
with condensed RVB gauge fields, with a possible weak ferromagnetism and
low temperature p-wave superconductivity are also suggested at higher
dopings. 

\end{abstract}



\begin{multicols}{2}[]

Recent discovery of superconductivity in \nxcob (\tc$\approx 5K$) 
by Takada and collaborators\cite{takada} marks a milestone in the 
search for new layered transition metal oxide 
superconductors. Following heels Wen et al.\cite{wensc}
 have reported superconductivity in \nycob ($A = Na,K$) with 
a high \tc $\approx 31~K$. In the same family of planar \cob based 
metals weak ferromagnetism\cite{fm} and high temperature curie 
susceptibility have been observed.  \ncob has been shown\cite{thPower}
 to be a very good 
metal with anomalously large thermoelectric power. It is becoming clear 
that strong electron correlation is at work, resulting in anomalous 
behavior and possible new electronic phases.

In this letter we model a reference \cob layer as an orbitally non degenerate
spin-\half antiferromagnetic Mott insulator on a triangular lattice.
Consequently \nxcob is described as an electron doped Mott insulator by 
a t-J model. We have performed an RVB mean field analysis of this model.
As there is a rich variety of possible phases, we do not go to the details
of the mean field theory but provide qualitative arguments (that go 
beyond mean field theory) for i) a reference chiral RVB state,
ii) a chiral RVB metallic (spin gap) state, iii) a weak ferromagnetic
state at higher doping and iv) PT violating d and p-wave 
superconductivity at low temperatures. 

\nxcob ($x\approx 0.35$ and $y \approx 1.3$) consists\cite{takada}
 of two dimensional
\cob layers separated by thick insulating layers of $Na^{2+}$ ions and
$H_2O$ molecules. \cob has the structure of layers in$CdI_2$.
It is a triangular net of edge sharing oxygen octahedra(figure 1); 
$Co$ atoms are at the center of the octahedra forming a 2D triangular 
lattice. Oxygen octahedra have a trigonal distortion - a stretch along 
a body diagonal direction 
of the embedding cube. For convenience we choose the corresponding body 
diagonal direction as Z-axis. The trigonal stretch makes the $O-Co-O$ 
angle $\approx 98^0$. 
A simple way to understand this structure is to imagine a triangular 
lattice on the XY plane with sub lattices A, B and C. Fill A and C 
sublattices with oxygen atoms and B sub lattice with Co atoms. Displace 
sublattice A and C layers on opposite directions along Z-axis 
by a same amount $\frac{z_0}{2}$. When $z_0 >\sqrt{\frac{3}{2}}a$ we 
get the 
desired structure with every Co atom surrounded by an octahedron with 
trigonal stretching; Here $a$ is the distance between neighboring Co 
atoms in the triangular lattice.  
\begin{figure}[h]
\epsfxsize 6.2cm
\centerline {\epsfbox{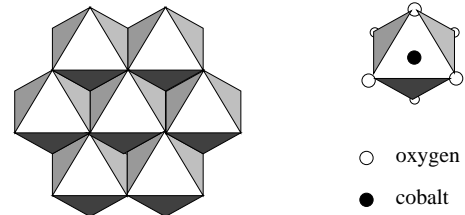}}
\caption{Structure of $CoO_2$ layer. A triangular network of edge 
sharing oxygen octahedra. Co atoms are at the center of the oxygen
octahedra.}
\end{figure}
Strong electron affinity of oxygen should lead to a complete electron 
transfer from $Na$ atoms of $Na_x.yH_2O$ layers in \nxcob 
and $Na$ or $K$ layers in \nycob, resulting in electron doped \cob layers. 

Experimentally \ncob is a strongly anisotropic\cite{susc} metal. 
The ab-plane resistivity is rather low with 
$ \rho \sim 10\mu \Omega$ at low temperatures and  
$ \rho \sim 200\mu \Omega$ at $300K$. C-axis resistivity is high
indicating some kind of confinement of charges, similar to the 
planar cuprates, at temperatures above $\sim 200 K$. It is also
interesting that such a good metal exhibits\cite{susc}  Curie-Weiss
$ \chi \sim \frac{C}{T - \Theta}$, rather than Pauli susceptibility 
at high temperatures; and $\Theta \approx - 118 K$. 
For $Na_{0.75}CoO_2$ weak ferromagnetism
has been observed\cite{fm} below about $22 K$. In another \cob layered 
compound called misfit layer compound, weak ferromagnetism has 
been reported\cite{fm}  below $3.5 K$. As mentioned earlier, two groups have
reported\cite{takada,wensc} superconductivity. 

In a neutral reference \cob layer the nominal valence of $Co$ atom is
$Co^{4+}$; i.e. a $3d^5$ ion. In an octahedral environment the 3d levels
are split (figure 2); The trigonal distortion of the oxygen 
octahedra causes further splitting of the 3d levels. The lower three 
fold degenerate $t_{2g}$ levels are split into a non-degenerate $d_{z^2}$ 
state with a doublet below, denoted as $e_g(t_{2g})$ in figure 2. 
The ground state 
configuration is an orbitally non-degenerate spin-\half, low spin state. 
In our coordinate system we choose the direction of trigonal distortion
to be the z-axis; so the top non-degenerate split orbital of the 
$t_{2g}$ manifold is a $3d_{z^2}$ orbital. Thus in the nominal
charge state $Co^{4+}$, we have an unpaired electron in the $d_{z^2}$
orbital, making $Co^{4+}$ ground state an orbitally non-degenerate
spin-\half state. In the same way for $Co^{3+}$ we have two electrons 
filling the $d_{z^2}$ level making the ground state an orbitally 
non-degenerate spin singlet. Our simple quantum chemical picture is
supported by various experiments including $Co$ NMR study\cite{susc}
 in \ncob.  It is interesting to note that our \cob layer
with strong trigonal distortion has escaped some important effects which 
should work against superconductivity, namely Jahn Teller distortion, 
Hund coupling and possible high spin ground states of $Co$ ions. 
\begin{figure}[h]
\epsfxsize 6cm
\centerline {\epsfbox{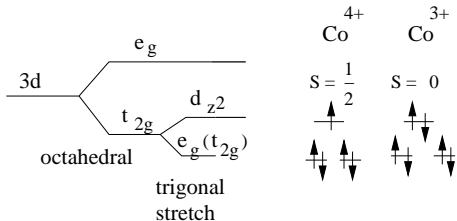}}
\caption{Crystal field split 3d levels of cobalt.} 
\end{figure}
As in other transition metal oxides, hybridization of $d_{z^2}$
with symmetry adapted oxygen orbitals and the strong Hubbard repulsion
in the $d_{z^2}$ should lead to the usual super exchange interaction 
between neighboring magnetic $Co^{4+}$ions. As the super exchange paths 
are not $180^0$ paths, the antiferromagnetic coupling will be reduced 
in strength. We make an estimate of the super exchange constant and 
parameters for our tight binding model using the electronic structure 
calculation of Singh\cite{singh} for \ncob. Singh finds an electron like 
Fermi surface, shown in figure 3. The states close to the Fermi level 
arise predominantly from the cobalt $3d_{z^2}$ orbitals; our quantum
chemical arguments are in agreement with Singh's result. The bilayer
type splitting found by Singh for \ncob is not important for us as 
our \cob layers are well insulated by the $Na_x.yH_20$ layers.
As a first approximation we ignore the small electron like pockets 
found by Singh. It is likely that electron correlation will push 
these minority bands away from the Fermi level.

\ncob contains equal number of $Co^{3+}$ and $Co^{4+}$ ions giving 
an average occupancy of $1.5$ electrons in the valence $d_{z^2}$
orbital. Thus the $d_{z^2}$ based single band is $\frac{3}{4}$ filled.
In a simple tight binding model keeping only the nearest neighbor 
hopping we get the following band dispersion:
\be
\epsilon_k = 
-2t \left( \cos k_x + 2 \cos {\frac{k_x}{2}} \cos {\sqrt{\frac{3}{2}}k_y}
\right)
\ee
We can estimate the value of the hopping parameter
$t$ by fitting to Singh's result. We get a value of $t \approx -0.1$,
corresponding to a band width of $\approx 1.0~eV$.
It is important to note that it is the negative sign of the hopping
parameter that makes the Fermi surface electron like for our $\frac{3}{4}$
filled band. If $t$ were negative we would have got hole like Fermi 
surfaces. In view of the strong particle hole asymmetry in a triangular
lattice the sign of $t$ is important.     
\begin{figure}[h]
\epsfxsize 3cm
\centerline {\epsfbox{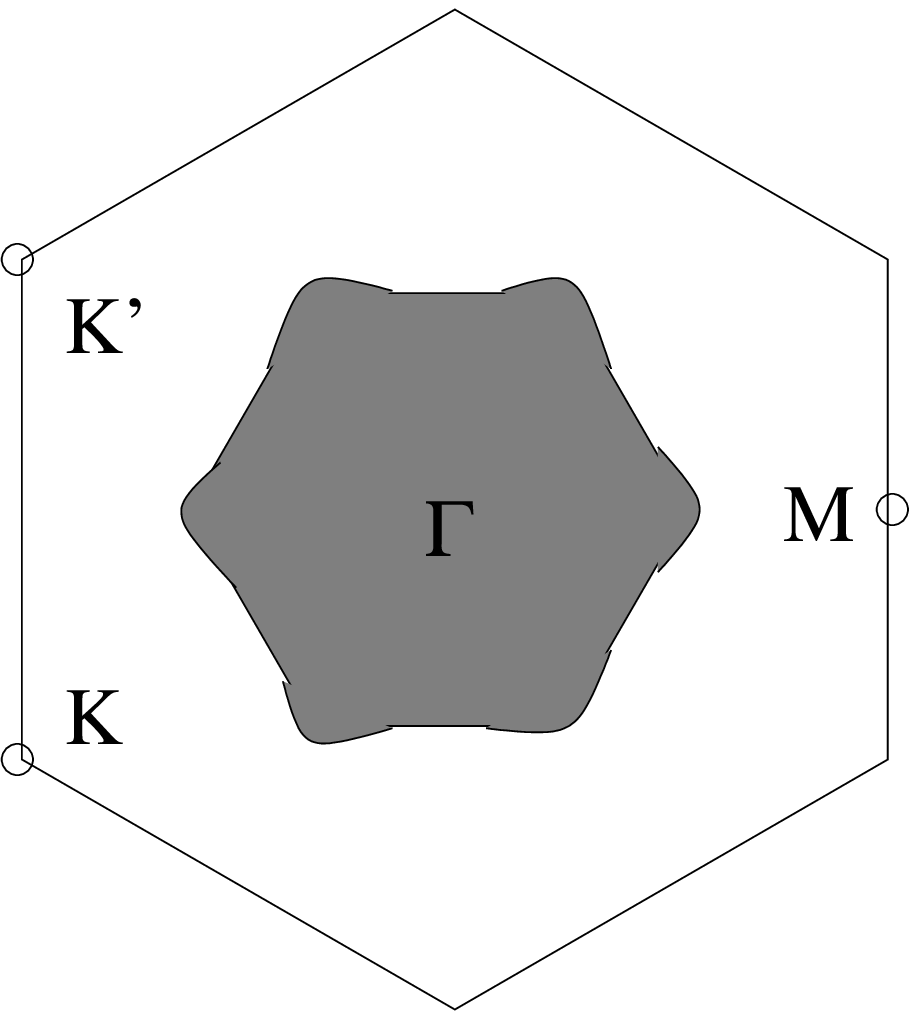}}
\caption{Fermi surface of electron doped $CoO_2$ layer, as calculated 
by Singh for \ncob, ignoring small Fermi surface pockets.}
Since coulomb interaction in the effective $3d_{z^2}$ orbital is 
$\sim 5~$to$~7~eV$, the net super exchange interaction between two 
neighboring $Co^{4+}$ ions is $J \equiv \frac{4t^2}{U} \sim 6~$
to$~8~meV$. Using the paramagnetic curie temperature $\Theta \approx 
-118~K$ obtained from  susceptibility measurements in \ncob we 
independently estimate $J \approx 7~meV$, assuming an average of 
3 nearest neighbor $Co^{4+}$ ion for a given $Co^{4+}$ ion. These 
considerations lead us to a t-J model for the electron doped \cob 
layer:

\bearr
H_{{\rm tJ}} =
 -t  \sum_{\langle ij\rangle} C^{\dagger}_{i\sigma} C^{}_{j\sigma}
+ h.c. +  J \sum_{\langle ij\rangle} 
({\bf S}_{i }\cdot{\bf S}_{j} - \frac{1}{4}n_i n_j) \nonumber
\eearr
with the local constraint $n_{i\uparrow} + n_{i\downarrow} \neq 0$. 

Existence of superconductivity in 2D t-J model in a square lattice
is no more doubted, thanks to RVB 
theory\cite{pwaScience,bza,gauge,krs,affleck} and related
recent\cite{nandini} variational and numerical efforts. 
The singlet proliferation tendency arising from 
the super exchange, contained in the J term seems to be sufficient 
to induce a robust spin singlet superconducting state\cite{vbam}.
We believe the same is true for the triangular 
lattice. However, the enhanced frustration could modify the symmetry 
of the superconducting state or introduce novel quantum states
such as chiral RVB metal with weak ferromagnetism.  
Further, possibility of superconductivity in a repulsive Hubbard model 
on a triangular lattice at and close to half filling has been studied 
by various authors\cite{dagotto,flex} invoking spin fluctuation mediated 
pairing, mainly in the context of organic superconductors. We take an RVB 
approach, as it is a natural way to study a system in its strong 
correlation limit and dominated by spin singlet correlation and chiral 
fluctuation. We study the undoped case first and discuss a simple 
RVB mean 

field theory for chiral RVB state. The doped Chiral RVB state is 
discussed next. 
As in the case of cuprates the RVB mean field solutions are guidelines 
to pick out the important phases and to map a phase diagram. It becomes 
quantitative and accurate when used in conjunction with Gutzwiller 
projection on the RVB mean field states.

Spin-\half Heisenberg Antiferromagnet on a triangular lattice has been
a guiding model for Anderson's RVB theory\cite{pwa73}, 
which took new meaning and new forms in the context of cuprates 
superconductors in the hands of Anderson, collaborators and others.
Kalmayer and Laughlin\cite{laughlin1} proposed a novel short range RVB 
wave function, called a chiral spin liquid state, that violated P \& T 
symmetry; it has non-zero expectation value of spin chirality:  
$\langle {\bf S}_{i} \cdot ({\bf S}_{j} 
\times {\bf S}_{k}) \rangle \neq 0 $. This state has certain deep
connection to fractional quantum Hall states. The energy of this state 
per spin is higher by about 9\%,
compared to better ground state energy estimates. 
Though not better in  energy, this above work
established the possibility of a chiral RVB state with a spin gap 
that also has a manifest quantum number fractionization through 
the existence of a well defined spinon excitation, that are anyons.
Consequently, when such a state is doped with holes for example, the 
holes undergo spin charge decoupling and the holons become anyons. 
This prompted Laughlin\cite{laughlin2,wwz} to argue for a powerful 
pairing correlation between holons arising from the novel exchange 
statistics. Later the anyon statistics was understood to have arisen 
from the attachment of appropriate RVB gauge field fluxes to particles.

We start with the slave boson representation for our t-J mode:
\be
H_{\rm tJ} = -t\sum_{\langle ij\rangle} 
s^\dagger_{i\sigma} s^{}_{j\sigma} + H.c.
-J\sum_{\langle ij\rangle} 
s^\dagger_{i\sigma} s^{}_{j\sigma} 
s^\dagger_{j\sigma'} s^{}_{i\sigma'} 
\ee
with the local constraint, $ d^\dagger_{i} d^{}_{i} +
\sum_{\sigma} s^\dagger_{i\sigma} s^{}_{i\sigma} = 1$. 
Here $s^{}_{i\sigma}$ are fermion operators for spin-\half
singly occupied states  and $d^{}_i$ are the bosonic operators for the 
doubly occupied spin singlet states. 

We first consider the undoped case, the spin-\half Heisenberg model on
a triangular lattice. Lee and Fang\cite{leeFeng} have 
performed RVB mean field analysis 
for this case. We briefly review their results. The mean field Hamiltonian
$H_{\rm mF} = -J\sum_{\langle ij\rangle} \chi_{ij} 
s^\dagger_{j\sigma} s^{}_{i\sigma} + ..$ is obtained by the factorization
$\tau_{ij}\tau_{ji} \rightarrow  \chi_{ij}\tau_{ji}$ etc., 
where $\tau_{ij} \equiv \sum_{\sigma} s^\dagger_{i\sigma} s^{}_{j\sigma}$.
The RVB order parameter $\chi_{ij} \equiv |\chi_0| e^{i\theta_{ij}}$
with $\theta_{ij} \equiv {\int}^{i}_{j} {\bf A}\cdot d{\bf l}$. Here ${\bf A}$
is the spatial component of the RVB gauge field\cite{gauge}. In the Lee-Feng 
mean field solution $\oint {\bf A}\cdot d{\bf l} = \pm 
\frac{\pi}{2}$ in every elementary triangle.

The single spinon dispersion acquires a gap at the fermi level and has
the form:
\bearr
\varepsilon_{k} = \pm J \alpha 
\left(\cos^2 k_x + 2 \cos^2 {\frac{k_x}{2}} \cos^2 {\sqrt{\frac{3}{2}}k_y}
\right)^{\frac{1}{2}}
\eearr
where $\alpha \approx 0.603 {\sqrt 3}$. The Chiral spin character
of Lee-Feng solution is obtained through the Wen-Wilczek-Zee identity\cite{wwz}
\be
{\bf S}_{i} \cdot ({\bf S}_{j} \times {\bf S}_{k}) \sim
i(\tau_{ij} \tau_{jk}\tau_{ki} - \tau_{ik} \tau_{kj}\tau_{ji})
\sim e^{i\oint {\bf A}\cdot d{\bf l}}
\ee
which connects the spatial component of the U(1) RVB gauge field with
three spin chirality. Thus the two degenerate solutions corresponding 
to uniform fluxes $\pm \frac{\pi}{2}$ through every elementary 
triangle give us the two degenerate PT violating chiral RVB states. 
Lee and Feng also found numerically that the energy of the above
mean field chiral RVB state, on Gutzwiller projection becomes very 
close to Kalmayer-Laughlin state.  We have also found a good overlap 
between Kalmayer-Laughlin wave function and Gutzwiller projected 
Lee-Feng RVB wave function using a procedure of 
Zou and Laughlin\cite{zouLaughlin}.

The $120^0$ 3-sublattice AFM order with a two fold planar chirality
degeneracy is to be viewed as obtained by condensing spinon pairs in
spin triplet state at the appropriate wave vector. As in cuprates,
even a small amount of doping, in view of the large dopant kinetic
energy removes the long range 3-sublattice order. We also find that
from our  RVB mean field analysis of the t-J model a locally stable 
chiral RVB state for a range of doping. {\em This means we can use the
$\frac{\pi}{2}$ flux RVB state as a reference state for a range of
doping in the metallic state}.

The situation is similar to cuprates, where Affleck-Marston's
$\pi$ flux RVB state\cite{affleck}, a state that respects PT symmetry, 
is useful to 
understand the spin gap phase. Absence of a sharp phase transition into 
the spin gap phase in the $x-T$ plane for cuprates seems to indicate 
that in the 
metallic spin gap phase the RVB flux is pinned to the PT symmetric 
value $\pi$. In our case, as our reference chiral RVB state has a
flux of $\frac{\pi}{2}$, a strongly PT violating value, we believe 
that our spin gap phase will be a chiral RVB metallic state.
It will be important to look for PT violation signals in experiments.

Experiments have shown weak ferromagnetism at high electron doping.
This may be explained as follows. Anderson has recently 
argued\cite{pwa2002} that the effect of dopant dynamics in cuprates 
is to produce local spin chirality and induced ferromagnetic interaction. 
We apply similar arguments for our case (figure 4) remembering 
that our
\begin{figure}[h]
\epsfxsize 7.5cm
\centerline {\epsfbox{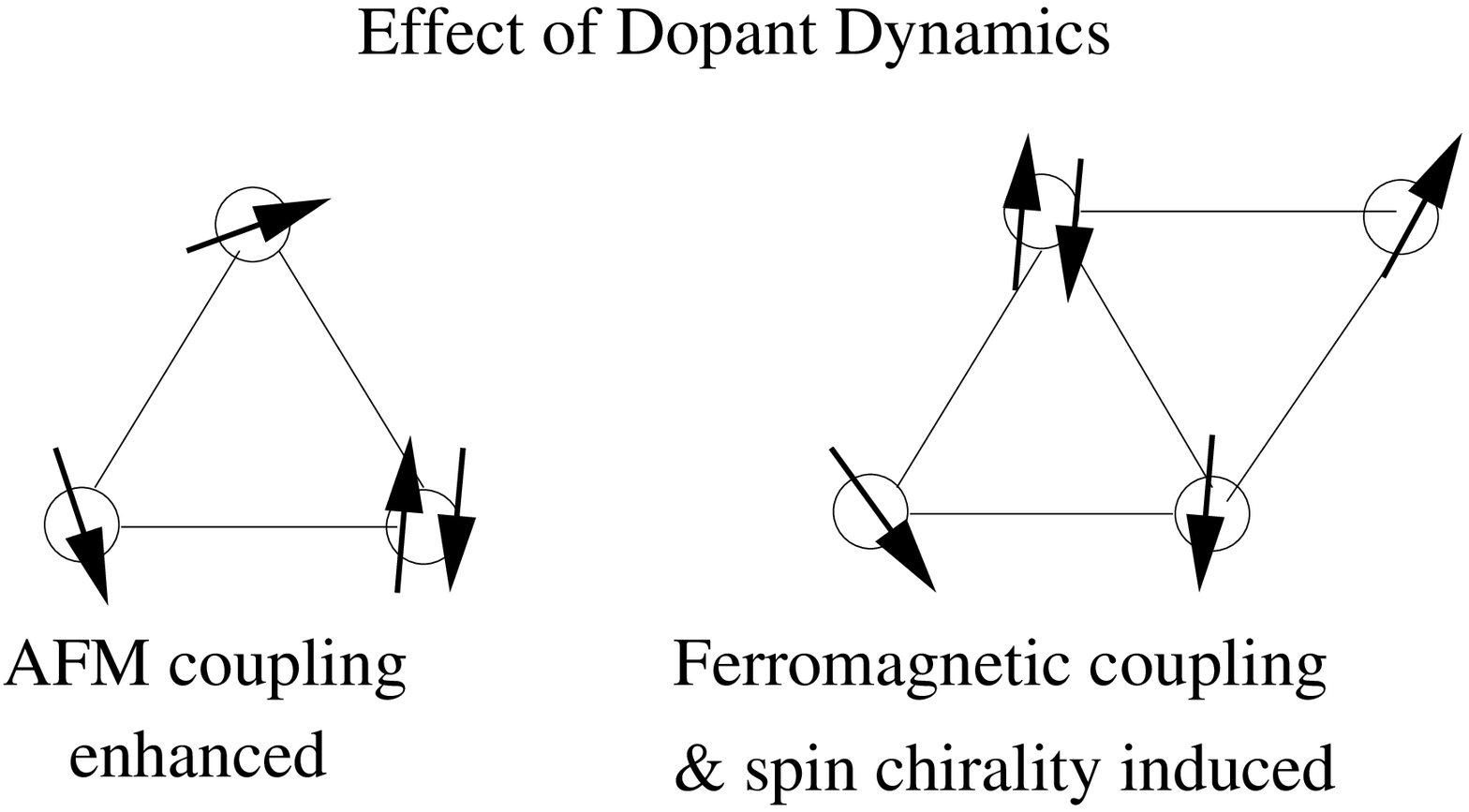}}
\caption{Effective spin-spin coupling induced by Dopant dynamics} 
\end{figure}
hopping integral has negative sign and our dopants are electrons.
Here a single extra electron 
performing a `closed loop hopping' in a triangle induces an extra 
antiferromagnetic coupling; this is because the above process permutes
an even number of spins. 

Thus we expect singlet stabilization
for a range of small doping. However, as the doping increases a 
carrier performing a closed loop hopping in a four spin cluster 
(figure 4) becomes important. As this process involves permutation
of an  odd number of spins, a ferromagnetic coupling is induced. 
Following Anderson we estimate this to be 
$J_{\rm eff} \approx J + xt$, for large $x$ ($t$ is -ve). 
In addition to the 
ferromagnetic coupling chirality is also favored by the above process.
Thus we believe that the weak ferromagnetism observed at high temperatures
is a chiral RVB metal with weak ferromagnetic moments induced in a 
novel chiral metal, where chiral fluctuations and ferromagnetism 
tendencies compete. Putting in the values of $J$ and $t$ it is easy 
to explain the range of ferromagnetic \tc $ \sim 3.5K ~{\rm to}~ 22K$ 
seen in experiments. 

Now let us discuss superconductivity. In the RVB mean field theory 
we write, following ref.8, the super exchange term as a BCS interaction, 
$ J \sum_{\langle ij\rangle} 
({\bf S}_{i }\cdot{\bf S}_{j} - 
\frac{1}{4}n_i n_j) \equiv  -J \sum_{ij} b^{\dagger}_{ij} b^{}_{ij}$
and perform the Bogoliubov-Hartree Fock factorization of the 
pairing term: $b^{\dagger}_{ij} b^{}_{ij} \rightarrow
b^{\dagger}_{ij} \Delta_{ij} + {\rm H.c.~ etc.}$, where 
$b^{}_{ij} \equiv \frac{1}{\sqrt 2} 
(s^{}_{i\uparrow} s^{}_{j\downarrow}- s^{}_{i\downarrow} 
s^{}_{j\uparrow})$ and $\Delta_{ij} \equiv \langle b^{}_{ij}\rangle$.
For cuprates, Kotliar\cite{affleck} found the 
$d_{x^2-y^2}$ wave state to have a lower energy. 
\begin{figure}[h]
\epsfxsize 8.8cm
\centerline {\epsfbox{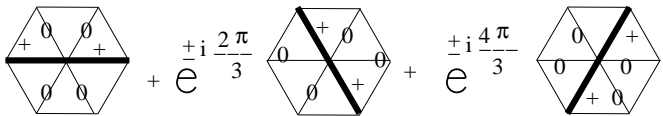}}
\caption{Superconducting order parameter
($|\Delta_{ij}| \neq 0$ on dark bonds) and relative phases in PT 
violating $d_1 \pm id_2$ states.}
\end{figure}
Triangular lattice the symmetry leads to two degenerate 
d-states $d_1$ and $d_2$, in the RVB mean field theory. 
For small doping PT vilating combinations $d_1 \pm id_2$ have 
lower energies. The order parameter pattern for the PT violating 
d-wave states are shown in figure 5. {\em Since the situation is more 
complex compared to cuprates, relative energetics of the 
extended-s, $d_1$, $d_2$ or the $d_1+id_2$, or the staggered or uniform
character of the spontaneously condensed RVB flux can be determined 
accurately only after studying the Gutzwiller projected mean 
field wave functions.}

Triplet superconductivity is also a distinct possibility as 
there is latent ferromagnetic tendency, arising from the dynamics of
the dopant charges. We will not go into the details of this.
All the above possibilities are summarized in a schematic phase
diagram in figure 6.  
\begin{figure}[h]
\epsfxsize 7.5cm
\centerline {\epsfbox{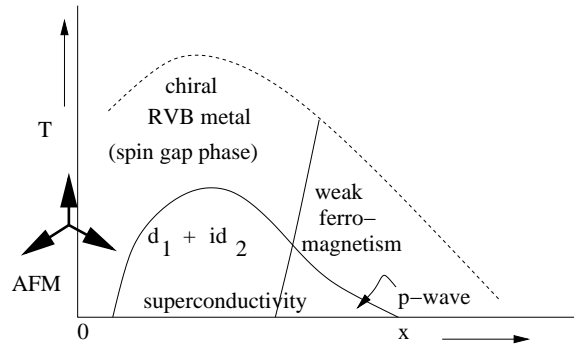}}
\caption{The schematic $x-T$ phase diagram.}
\end{figure}
As far as the scale of superconducting \tc is concerned, because 
of the small value of $J$, in our estimates we do not get \tc's far 
exceeding $30K$. We hope to present our quantitative analysis for 
the various phases discussed above in a future publication.

We conclude by stating that \cob based metals seems to be a new
class of strongly correlated systems that stabilizes novel resonating
valence bond states, that was not realized in cuprates. 


\end{figure}
\end{multicols}
\end{document}